\documentstyle[preprint,revtex,eqsecnum]{aps} \draft
\begin{document}
\newcommand{\lra}{\longrightarrow}
\newcommand{\ra}{\rightarrow}
\newcommand{\nc}{\newcommand}
\nc{\ba}{\begin{array}}
\nc{\ea}{\end{array}}
\nc{\acom}{\renewcommand{\arraystretch}{0.5}}
\nc{\astr}{\renewcommand{\arraystretch}{1}}
\nc{\sst}{\scriptstyle}
\nc{\ts}{\textstyle}
\nc{\ti}{\tilde}
\pagestyle{headings}

\begin{title}
ON THE QUANTIZATION OF THE N=2 SUPERSYMMETRIC \\ NON LINEAR SIGMA MODEL
\end{title}
\author{G. Aldazabal and  J. M. Maldacena }
\begin{instit}
Centro At\'omico Bariloche, 8400 Bariloche, \\
Comisi\'on Nacional de Energ\'{\i}a At\'omica, \\
Consejo Nacional de Investigaciones Cient\'{\i}ficas y T\'ecnicas, \\
and Instituto Balseiro, Universidad Nacional de Cuyo - Argentina
\end{instit}
\begin{abstract}
A method for quantizing the bidimensional N=2 supersymmetric non-linear sigma
model is developed. This method is both covariant under coordinate
 transformations (concerning the order relevant for calculation) and
explicitly N=2 supersymmetric.
The operator product expansion of the supercurrent
is computed accordingly, including also the
dilaton. By imposing the N=2 superconformal algebra  the equations for the
metric and
the dilaton are obtained. In particular, they imply that the dilaton is a
constant.
\end{abstract}

\newpage

\section{Introduction}

The N=2 supersymmetric sigma model appears in string theory in two different
contexts.
One is the propagation of a N=2 string in a curved background $^{\cite{vafa}}$.
The
other is  N=1 string compactifications,  since in this case N=1 space-time
supersymmetry requires a N=2 superconformal theory on the
compactified manifold $^{\cite{todosn=2}}$.  \\
A sigma model has two supersymmetries if and only if it is defined on a
K\"ahler
Manifold $^{\cite{alvgaumen=2}}$.              \\
It is well known that quantum corrections may destroy conformal invariance.
Thus,
further conditions on the manifold result from requiring conformal invariance
in the quantum theory.  \\
In order to compute  quantum corrections, it is convenient to maintain
the symmetries of the theory throughout the calculations. In this case,
the relevant symmetries are: target space reparametrization invariance and
worldsheet
N=2 supersymmetry. The latter
can be kept explicit by using the N=2 superfield formalism
$^{\cite{zumino}}$.\\
Reparametization
invariance is usually made explicit by expanding the fields in terms of normal
 coordinates around a classical background
$^{\cite{honerkamp,alvgaumecoord}}$.
Though this method works for  N=0,1, it cannot be extended
straightforwardly to  N=2. The reason for this is that the
superfield formalism treats holomorphic and antiholomorphic target
space coordinates differently, whereas the geodesic equation mixes them.
A previous attempt to overcome this
problem involved the use of prepotentials in order to solve the superfield
chirality constraints $^{\cite{howe}}$.       \\
Here we propose a method based on the modification of the geodesic
equation so that it does not mix holomorphic and antiholomorphic
coordinates. The chirality constraints are accomplished through Lagrange
multipliers.             \\
We use this method to derive the conformal anomaly.
 Instead of going through the computation of the $\beta$ function
$^{\cite{grisaru,callanbackground}}$ we choose the alternative of calculating
the generator algebra in the quantum theory.
We obtain the one loop equations
for the metric and the dilaton
by identifying the
symmetry breaking terms $^{\cite{banks,gerardo,japoneses}}$
and by setting them to zero.
The paper is organized as follows: in Section 2 we develop the method, in
Section 3 it is applied to calculate the operator product expansion
of the supercurrent and the equations for the metric and the dilaton are
obtained.
Conclusions are presented in Section 4.

\section{Quantization method}

An N=2 sigma model must be defined on a K\"ahler manifold
$^{\cite{alvgaumen=2}}$.
On such a manifold it is possible to choose complex coordinates
$(\phi^\mu , \ti{\phi}^{\ti{\nu}})$
so that the metric
is derived from a locally defined potential $K(\phi,\ti{\phi})$ in
the following way
\begin{equation} g_{\mu\nu}=0~~~~~~~g_{\ti{\mu}{\ti{\nu}}}=0 ~~~~~~~~~~
g_{\mu{\ti{\nu}}}=K,_{\mu{\ti{\nu}}}   \label{} \end{equation}
The superfield formulation  $^{\cite{zumino}}$ uses  the N=2 superspace
spanned by the coordinates
 (we work in Euclidean space)
${\cal Z}=({\rm z},\theta_{\rm z},\ti{\theta}_{\rm z},\bar{{\rm z}},
\theta_{\bar{{\rm z}}},\ti{\theta}_{\bar{{\rm z}}})$.
Complex conjugation on the
worldsheet (denoted with a bar) and complex conjugation in space-time
 (denoted with tilde) are distinguished, since they
correspond to two different symmetries.
The coordinates $\phi^\mu$, $\ti{\phi}^{\ti{\nu}}$ are superfields  and
obey the chirality constraints
\begin{equation} \ti{D}_a\phi^\mu = 0~~~~~~~~~~~~~~~~~D_a\ti{\phi}^{\ti{\nu}}=0
 \label{chicond} \end{equation}
where the covariant derivatives
\begin{equation} D_{a} = \frac{\partial}{\partial \ti{\theta}^{a}} - (\not \!
\partial\theta)_{a}~~~~~~~~~~~~~~~~
\ti{D}_{a} = \frac{\partial}{\partial \theta^{a}} - (\not \!
\partial\ti{\theta})_{a}
 \label{} \end{equation}
satisfy the conmutation relations
\begin{equation} \{D_a,D_b\}=0~~~~~~\{\ti{D}_a,\ti{D}_b\}=0~~~~~~~~~
\{ D_{a}, \ti{D}^{b}\} = 2 \not \! \partial_{a}^{~b}
 \label{conmrel} \end{equation}
In terms of the K\"ahler potential the action reads
\begin{equation} S=\frac{-1}{4\pi\alpha'} \int d^6 {\cal Z} K(\phi,\ti{\phi})
 \label{acck} \end{equation}
The chirality constraints are included in the action by means of Lagrange
multipliers (which are spinorial superfields), in the following way
\begin{equation}  S_{mult}=\frac{-1}{4\pi\alpha'} \int d^6{\cal Z}~(
\ti{\lambda}_\mu
\ti{D}\phi^\mu + \lambda_{\ti{\nu}}D\ti{\phi}^{\ti{\nu}} )
 \label{accmult} \end{equation}
The action (\ref{acck}) is invariant under the holomorphic coordinate
reparametrizations
\begin{equation} \ba{l} \phi^\mu \lra \phi'^\mu = f^\mu (\phi) \\
 \ti{\phi}^{\ti{\nu}} \lra \ti{\phi}'^{\ti{\nu}} = \ti{f}^{\ti{\nu}}(\ti{\phi})
\\ K(\phi,\ti{\phi}) \lra K'(\phi',\ti{\phi}') =  K(\phi,\ti{\phi})
\ea  \label{rephol} \end{equation}
The background field method consists in calculating quantum corrections around
an arbitrary solution $\phi_0^\mu$ for the classical equations of motion. The
field is then expressed as $\phi^\mu=\phi^\mu_0 + \Delta\phi^\mu$, where
$\Delta\phi^\mu$ is the quantum variable.
However, this decomposition is not covariant because $\Delta\phi^\mu$ is not
 a vector. \\
Since the
various terms of the expansion are evaluated on $\phi_0$, a convenient
quantum variable is a vector on the tangent bundle at $\phi_0$. In the
normal coordinate method $^{\cite{honerkamp,alvgaumecoord}}$ this vector is
the tangent vector at $\phi_0$ to the
geodesic that joins $\phi_0$ with $\phi$.
This expansion is very useful indeed for computations in  N=0 and
 N=1.
Due to the existence of the chirality constraints,
this is not the case when there is  N=2 supersymmetry.
Since the geodesic
equation mixes holomorphic and antiholomorphic coordinates, these constraints
become cumbersome when written in terms of normal coordinates.\\
The crucial observation is that
an equation  which is covariant only under holomorphic coordinate
transformations is needed. In fact, covariance under general reparametrizations
was lost when complex coordinates were chosen.\\
The modified equation reads
 \begin{equation} \ddot{\phi}^\mu + \Gamma^\mu_{\ \rho\delta}(\phi,\ti{\phi}_0)
\dot{\phi}^\rho \dot{\phi}^\delta =0 ~~~~~~~~
\ddot{\ti{\phi}^{\ti{\nu}}} +
\Gamma^{\ti{\nu}}_{\ \ti{\rho}\ti{\epsilon}}(\phi_0 , \ti{\phi})
\dot{\ti{\phi}^{\ti{\rho}}} \dot{\ti{\phi}^{\ti{\epsilon}}} =0
 \label{ecgeo} \end{equation}
The only difference it presents with the usual geodesic equation is that
the complex
conjugate coordinate is
kept fixed.
Recall that connections with mixed indices vanish in K\"ahler manifolds.
(\ref{ecgeo}) may be interpreted as the equations for the two geodesics that
join $(\phi_0,\ti{\phi}_0)$ with $(\phi,\ti{\phi}_0)$ and
$(\phi_0,\ti{\phi}_0)$ with $(\phi_0,\ti{\phi})$.
The quantum variables are
\begin{equation} \xi^\mu \equiv \dot{\phi}^\mu(t=0)~~~~~~~~~~~~~
\ti{\xi}^{\ti{\nu}} \equiv \dot{\ti{\phi}^{\ti{\nu}}}(t=0)  \label{}
\end{equation}
It is easy to see that $\phi$ ($\ti{\phi}$) depends only on
$\xi$ ($\ti{\xi}$).
 This does not mean that $\xi$
is a chiral field. In fact,  the passage to normal coordinates is not a mere
change of coordinates since it  depends on the point
$\phi_0({\cal Z})$ which
in turn depends on the world sheet coordinates. This may be expressed as
$\phi=\phi(\phi_0,\ti{\phi}_0,\xi)$.                    \\
Under this holomorphic normal coordinate expansion the chirality
constraint (\ref{chicond}) reads
\begin{equation} \ti{D}_a\phi=0 \lra \ti{D}_a\xi
 - \left( \frac{1}{2} R^\mu_{\ \delta_1 {\ti{\nu}}
\delta_2} \xi^{\delta_1} \xi^{\delta_2} + \frac{1}{3!} \nabla_{\delta_1}
R^\mu_{\ \delta_2 {\ti{\nu}} \delta_3} \xi^{\delta_1} \xi^{\delta_2}
\xi^{\delta_3} + \cdots \right)\ti{D}_a\ti{\phi}^{\ti{\nu}}_0 =0
 \label{} \end{equation}
and a similar expression is obtained for $D\ti{\phi}$ by just
replacing
$\xi$ by $\ti{\xi}$. From now on  we will concentrate on
holomorphic
quantities, antiholomorphic ones can be obtained by complex conjugating the
expressions.
The Lagrange multipliers are expanded in a
similar fashion. By noting that they are vectors under (\ref{rephol}),
their equation reads
\begin{equation} \nabla^M_t\nabla^M_t\ti{\lambda}_\mu = 0 ~~~~~\mbox{with}~~
\left(\nabla^M_t \right)^{\ \mu}_{\rho} = \delta^\mu_\rho \frac{d}{dt} -
 \Gamma^\mu_{\ \rho\delta}(\phi,\ti{\phi}_0) \dot{\phi}^\delta
		  \label{} \end{equation}
and $\ti{\chi}_\mu \equiv \nabla^M_t\ti{\lambda}_\mu(t=0)$  is the quantum
variable.\\
The solution to this equation can be written in terms of
$\phi(\xi,\phi_0,\ti{\phi}_0)$ as
\begin{equation} \ti{\lambda}_\mu =
\frac{\partial \phi^\rho}
{\partial \xi^\mu} (\ti{\chi}_\rho + \ti{\lambda}^0_\rho)
 \label{chil} \end{equation}
where $\ti{\lambda}^0_\rho$ is the classical solution.
The total action is written as
\begin{equation}
 S = S_0 + S_K +S_H + S_{mult}   \label{}
\end{equation}
where $S_0$ is the action evaluated on the classical solution. $S_K$ includes
those terms
coming from the K\"ahler potential expansion that contain both $\xi$ and
$\ti{\xi}$
\begin{equation}
    S_K = \frac{-1}{4\pi\alpha'} \int
    K,_{\mu{\ti{\nu}}}\xi^\mu\xi^{\ti{\nu}} +
    \frac{1}{4} R_{\mu_1{\ti{\nu}}_1
    \mu_2{\ti{\nu}}_2} \xi^{\mu_1} \xi^{\mu_2} \ti{\xi}^{{\ti{\nu}}_1}
    \ti{\xi}^{{\ti{\nu}}_2} + \cdots  \label{enserio}
\end{equation}
$S_H$
includes the terms which contain only holomorphic or antiholomorphic quantum
fields
\begin{equation}
S_H = \frac{-1}{4\pi\alpha'} \int\sum^{\infty}_{n=2} \frac{1}{n!}
T_{\mu_1 \cdots \mu_n} \xi^{\mu_1} \ldots
\xi^{\mu_n} + c.c.
\label{SH}  \end{equation}
$$\mbox{with}~~ T_{\mu_1 \cdots \mu_n} = \nabla_{\mu_1} \ldots \nabla_{\mu_n}
K - \ti{\lambda}^0_\rho \ti{D} \ti{\phi}_0^{\ti{\nu}}
\nabla_{\mu_1} \ldots \nabla_{\mu_{n-2}} R^\rho_{\ \mu_{n-1} {\ti{\nu}} \mu_n}
$$
and $S_{mult}$ is the part of (\ref{accmult}) that contains the quantum
multiplier $\ti{\chi}$
\begin{equation} S_{mult} = \frac{-1}{4\pi\alpha'}\int \ti{\chi}_\mu \ti{D}
\xi^\mu - \frac{1}{2} \ti{\chi}_\mu R^\mu_{\ \delta_1
{\ti{\nu}} \delta_2} \xi^{\delta_1} \xi^{\delta_2}
\ti{D} \ti{\phi}_0^{\ti{\nu}}  + \cdots   + c.c.
 \label{smult} \end{equation}
We  eliminate the metric $K_{\mu{\ti{\nu}}}(\phi,\ti{\phi}_0)$ from
(\ref{enserio}) by passing to the local ortonormal
frame $e^r_\mu(\phi_0,\ti{\phi}_0)$
and by referring all tensors to this frame.
\begin{equation}
e^r_\mu e^{\ti{s}}_{\ti{\nu}} K^{\mu{\ti{\nu}}}(\phi_0, \ti{\phi}_0) =
\delta^{r\ti{s}}~~~~~~ E^{\ti{r} \mu} e^r_\rho = \delta^\mu_\rho~~~~~~
 \xi^r = e^r_\mu \xi^\mu ~~~~~~\ti{\chi}^{\ti{s}} =
E^{\ti{s}\mu} \ti{\chi}_\mu  \label{vielbein} \end{equation}
The derivative $\ti{D}\xi$  gives rise to a connection
$\omega^{r\ti{s}}_{\ti{D}}$
\begin{equation}
(\hat{\ti{D}})^{r\ti{s}}  = \delta^{r\ti{s}} \ti{D} + \omega^{r\ti{s}}
_{\ti{D}}
{}~~~~~\mbox{where}~~
\omega^{r\ti{s}}_{\ti{D}} = e^r_\rho E^{\ti{s}\rho}_{'{\ti{\nu}}} \ti{D}
\ti{\phi}_0^{\ti{\nu}}
 \label{} \end{equation}
The quadratic form appearing in the action can be rewritten as
\begin{equation} S^{(2)} = \frac{-1}{4\pi\alpha'}
\int \ti{\xi}^{r} \xi^r + \ti{\chi}^{\ti{r}}\hat{\ti{D}} \xi^r
+ \chi^r \hat{D} \ti{\xi}^{\ti{r}}
 \label{s2} \end{equation}
and it is not invertible. As usual, this is
due to a gauge symmetry of the original action. This symmetry operation, which
 affects only the Lagrange multipliers, is
\begin{equation} \ti{\lambda}_\mu \lra \ti{\lambda}_\mu + P \ti{\epsilon}_\mu
 \label{gsim} \end{equation}
where we defined the projector
\begin{equation}
 P_{a}^{~b} =-\frac{\ti{D}\not \! \partial D}{2\partial^2}\delta^b_a
+\frac{1}{2}\frac{\ti{D}^2 D^2}{16 \partial^2} \delta^b_a
+ \frac{\ti{D}_a(D\not \! \partial)^b}{4\partial^2}
 \label{p} \end{equation}
Therefore only part of the superfield $\ti{\lambda}$ (that in the Kernel of P
(\ref{p}) ) is necessary
in order to
fix the chirality condition (\ref{chicond}). This is due to the fact that the
 $\ti{D}_a, D_a$ operators
are not invertible, as results from the conmutation relations (\ref{conmrel}).
In terms
of the  quantum variable $\ti{\chi}^{\ti{r}}$ (\ref{chil}, \ref{vielbein})
the  symmetry (\ref{gsim}) reads
\begin{equation}
\ti{\chi}^{\ti{r}} \lra \ti{\chi}^{\ti{r}} + E^{\ti{r}\rho}
\frac{  \partial \phi^\mu  }
{ \partial \xi^\rho } P\ti{\epsilon}_\mu
 \label{calchi} \end{equation}

When a gauge fixing condition is chosen, the associated Fadeev-Popov
determinant  also has a gauge symmetry and so on
ad infinitum.
The same phenomenon appeared in the
previous covariant quantization scheme $^{\cite{howe}}$.
Such a process can be ended   at some stage by fixing the gauge in a non
covariant fashion.
We choose the non covariant condition
\begin{equation} P\ti{\chi}=0  \label{gcond} \end{equation}
The ghost action is
obtained
by performing a gauge transformation of the gauge fixing condition in the usual
way
\begin{equation}
b^{r} P \delta \ti{\chi}^{\ti{r}} =
b^{r} P  E^{\ti{r}{\rho}} \frac{\partial \phi^\mu}
{\partial \xi^\rho} P c_\mu
 \equiv {\cal L}_{gh}(b,c)  ~~~~~~
S_{gh} = \int d^6 {\cal Z}~({\cal L}_{gh} +\ti{\cal L}_{gh})
 \label{lfant} \end{equation}
However, as the symmetry (\ref{calchi}) has a fermionic parameter, the ghosts
are bosonic
superfields. Note that only $\xi$
($\ti{\xi}$) appears in ${\cal L}_{gh}$ ($\ti{\cal L}_{gh}$).
As  we said above, the ghost action has a new gauge symmetry because
$P$, being a projector,  has zero modes. We fix it with the conditions
\begin{equation} Sc_\mu = 0 ~~~~~~~~~~~~~~~~~~~S^{\dagger}b^r = 0
 \label{} \end{equation}
where $S=1-P$ and $S^{\dagger}$  is the operator obtained by integrating
the
operator $S$ by parts. The new Fadeev-Popov determinant decouples because the
gauge condition (\ref{gcond}) is not covariant. We will see, however, that
for our
calculation the ghosts do not contribute.

The reason why we cannot find a fully
explicitly covariant quantization may be understood from the fact that the
one loop counterterm:
$\delta K \sim \frac{1}{\epsilon}\log\det(K,_{\mu{\ti{\nu}}})$ is
not a scalar under holomorphic
reparametrizations. Of course, once integrated over the whole superspace
an invariant
quantity is obtained. If the theory were completely covariant this counterterm
could not possibly appear. In this approach, this divergence emerges from the
non-covariant ghost action, in particular, from the superdeterminant
coming from its quadratic part.

We take as the free action the part of (\ref{s2}) which does not contain
the connection.
This separate treatment of the connection produces a new source of
non covariance, which is also present in the formalism used for the cases
N=0,1. The propagators for the fields are
\begin{equation}
\langle \xi^r(z) \ti{\xi}^{\ti{s}}(z')\rangle =
-4\pi\alpha' \frac{\ti{D}^2 D^2}{16 \partial^2}  \delta^{r\ti{s}}
 \label{proppsi} \end{equation}
\begin{equation}
\langle \ti{\chi}^{\ti{r}}_a(z) \xi^s(z')\rangle  =
-4\pi\alpha' \left( -\frac{(\not \! \partial D)_a}{2\partial^2} +
\frac{\ti{D}_a D^2}{16 \partial^2} \right) \delta^{\ti{r}s}   \label{proplpsi}
 \end{equation}
\begin{equation}
\langle \chi^r_a(z) \ti{\chi}^{\ti{s}\,b}(z')\rangle =
-4\pi\alpha' \left[ \frac{\not \!  \partial_a^{~b}}{2\partial^2} \left(
-\frac{\ti{D}^2D^2}{16\partial^2} \right) + \frac{D_a\ti{D}^b}{16\partial^2}
+ \frac{1}{8} \frac{D^2\ti{D}^2}{16\partial^2}\frac{\not \! \partial_a^{~b}}
{\partial^2} \right] \delta^{r\ti{s}}
 \label{propll} \end{equation}
Note that the propagators involving $\ti{\chi}, \chi$  have a softer
ultraviolet behavior.
In fact we will see that power counting arguments can be used to discard many
terms from the expansion.

\section{Calculation of the conformal anomaly}

Based on  ideas of Banks et al $^{\cite{banks,gerardo,japoneses}}$,
we calculate the conformal anomaly
by evaluating the generator algebra in the quantum theory. This algebra
can be extracted from
 the divergent terms when ${\cal{Z}} \ra {\cal{Z}'}$ in
the supercurrent operator product expansion.
The calculation is accomplished perturbatively in $\alpha'$.

The N=2 superconformal algebra is the direct sum of  holomorphic and
antiholomorphic sectors. From now on we will consider only the holomorphic
sector, which is  described by  the supercurrent
\begin{equation}
J_{\rm z} = j_{\rm z} + \ti{\theta}_{\rm z} S_{{\rm z}{\rm z}} -
\theta_{\rm z} \ti{S}_{{\rm z}{\rm z}} +
\theta_{\rm z}\ti{\theta}_{\rm z} T_{{\rm z}{\rm z}}  \label{} \end{equation}
where $j_{\rm z}$ is the generator of $U(1)$ transformations,
$S_{{\rm z}{\rm z}}$ and
$\ti{S}_{{\rm z}{\rm z}}$   are
associated with the two supersymmetries and $T_{{\rm z}{\rm z}}$ is the
stress-energy tensor.
In terms of the operator product expansion the algebra reads
$^{\cite{divecchia}}$
\begin{equation} J_{\rm z}({\cal Z})J_{\rm z}({\cal Z}') \sim
\frac{\frac{4}{3}c}{(\Delta {\cal Z})^2} +
\frac{4 \ti{\delta}_{\rm z}\delta_{\rm z}}{(\Delta {\cal Z})^2}
J_{\rm z}({\cal Z}^c) +
\frac{2}{\Delta {\cal Z}}
(\ti{\delta}_{\rm z}D_{\rm z} - \delta_{\rm z} \ti{D}_{\rm z})
J_{\rm z}({\cal Z}^c)
 \label{opej+} \end{equation}
where
$$
\Delta {\cal Z} = {\rm z}-{\rm z}' -
\theta_{\rm z}\ti{\theta}'_{\rm z} - \ti{\theta}_{\rm z}\theta'_{\rm z}
{}~~~~~~~~~~~~~
\delta_{\rm z} = \theta_{\rm z} - \theta'_{\rm z} ~~~~~~~~~~~~~
\ti{\delta}_{\rm z} =
\ti{\theta}_{\rm z} - \ti{\theta}'_{\rm z}   $$
$$ {\cal Z}^c= \frac{{\cal Z}+{\cal Z}'}{2}=
\left( \frac{{\rm z}+{\rm z}'}{2}, \frac{\theta_{\rm z} + \theta'_{\rm z}}{2},
\frac{\ti{\theta}_{\rm z} +  \ti{\theta}'_{\rm z}}{2} \right) ~~~~~~~~~
c=\mbox{central charge}  $$
In our model the supercurrent is
\begin{equation} J_{\rm z} =
 \frac{-2}{\alpha'} \ti{D}_{\rm z} \ti{\phi}^{\ti{\nu}}  D_{\rm z} \phi^\mu
K,_{{\ti{\nu}}\mu}(\phi,\ti{\phi})  \label{} \end{equation}
The dilaton field can be included by adding a new term to the current
\begin{equation}
 J^{dil}_{\rm z} = 2 [ \ti{D}_{\rm z}, D_{\rm z} ] \Phi(\phi,\ti{\phi})
 \label{jdil} \end{equation}
The situation is similar in the N=0,1 cases where the dilaton
disappears from the action once the world sheet gravity or supergravity
 is gauge fixed, but it still appears in the stress energy tensor.

We calculate the OPE up to one loop for the less divergent terms and up to
two loops for the central charge term. As (\ref{jdil}) is one order
higher in $\alpha'$,
dilaton graphs need less loops for a given order in $\alpha'$. Only
those terms which are proportional to $\theta_{\rm z},
\ti{\theta}_{\rm z}, \theta'_{\rm z},
\ti{\theta}'_{\rm z} $ need to be considered
because other terms (e.g. one with $\theta_{\bar{{\rm z}}}$) are
 related to the operator product of the auxiliary components of the
supercurrent.
Dimensional regularization is used to deal with UV divergences. IR
divergences
are regulated with a mass cutoff and,
 though present in the propagators,
disappear from the final result.

Keeping in mind that  only terms divergent when
${\cal{Z}} \ra {\cal{Z}'}$ are needed,   power counting arguments are useful
in order to  discard some
terms.  The superspace volume element $d^6{\cal Z}$ and
the propagator
$\langle \xi \ti{\xi} \rangle$   have  dimension zero.
The derivatives
$D_a$, $\ti{D}_a$ have dimension $\frac{1}{2}$. The supercurrent has
dimension one
and only positive dimension terms are to be evaluated. If a derivative acts on
a background field, then the dimension of the $\Delta{\cal Z}$ dependent part
gets
reduced. This occurs when  a vertex coming from (\ref{SH})
with all incoming
(or all outgoing) $\langle \xi \ti{\xi} \rangle$ propagators is present.
One
of the $\ti{D}^2$ acting on the
$\langle \xi \ti{\xi} \rangle$ lines can be integrated by parts and applied
to the tensor evaluated in
$\phi_0$, reducing  the effective dimension by one.
This is a highly desirable circumstance because
those terms are not invariant under a K\"ahler gauge transformation, i.e.
 $K \ra K + f(\phi) + \ti{f}(\ti{\phi})$ under which the metric
is invariant.
A similar argument is applied for the ghost loops with more than one
incoming line because they also have  only incoming or only outgoing
$\langle \xi \ti{\xi} \rangle $
lines.
Similarly, each $\langle \xi \ti{\chi} \rangle$  propagator decreases the
dimension in $\frac{1}{2}$ and
each $\langle \chi \ti{\chi} \rangle$ line in 1. Therefore no $\langle \chi
\ti{\chi} \rangle$ will appear.

The order zero in $\alpha'$ graphs contributing to $\langle J_{\rm z}({\cal Z})
J_{\rm z}({\cal Z'}) \rangle $ are shown in figure 1.
In computing mean values nothing is lost because $\phi_0$ is an
arbitrary classical solution.

In order to extract the symmetry breaking terms, the right hand side
of (\ref{opej+}) is substracted.
Therefore it is necessary to compute $\langle J \rangle $
via the graphs shown in
figure~2.

Some graphs are UV divergent but they cancel when the above mentioned
difference
is performed. Divergences should in fact be eliminated considering the
contribution
of the one loop counterterm since this is a one loop calculation. It can be
verified, however, that the contribution of the counterterm
also cancels when the  difference is performed.
 The results of the graphs that involve the noncovariant
connection $\omega^{r\ti{s}}$ must be recovariantized.
These would be indeed covariant if the
graphs
with two connection insertions had been included.
In the cases N=0,1 these noncovariant
terms are also present but, contrary to this case, they cancel among
themselves.
In those cases the action has a quadratic term proportional to the
curvature that is here instead hidden in the connection.

The central charge term is computed up to first order in $\alpha'$  (two
loops), the
graphs are shown in figure 3.

Adding these results we obtain
$$ \langle J_{\rm z}({\cal Z})J_{\rm z}({\cal Z}')\rangle
- \frac{4\ti{\delta}_{\rm z}\delta_{\rm z}}
{(\Delta{\cal Z})^2} \langle J_{\rm z}({\cal Z}^c)\rangle  -
\frac{2}{\Delta{\cal Z}}
(\ti{\delta_{\rm z}} D_{\rm z} -
\delta_{\rm z} \ti{D}_{\rm z} ) \langle J_{\rm z}({\cal Z}^c)\rangle  \sim
$$
\begin{equation} \sim
\frac{4D -8\alpha'(-\frac{1}{4} R + \nabla^\mu \nabla_\mu \Phi -
\nabla^\mu \Phi \nabla_\mu \Phi)}{(\Delta{\cal Z})^2}+
 \end{equation}
      $$
+\frac{2\Delta \bar{\cal Z}}
{(\Delta{\cal Z})^2}
\left\{
\ti{\delta}_{\rm z} D_{\bar{{\rm z}}}[(R_{{\ti{\nu}}\mu} -
2 \Phi,_{{\ti{\nu}}\mu})
\ti{D}_{\bar{{\rm z}}}\ti{\phi}^{\ti{\nu}}_0 D_{\rm z}\phi^\mu_0]
-
\delta_{\rm z} \ti{D}_{\bar{{\rm z}}}[(R_{{\ti{\nu}}\mu}-
2\Phi,_{{\ti{\nu}}\mu})
\ti{D}_{\rm z}\ti{\phi}^{\ti{\nu}}_0 D_{\bar{{\rm z}}}\phi^\mu_0]
\right.
$$ $$
\left.
-2 \delta_{\rm z}D_{\bar{{\rm z}}}(\nabla_{\ti{\nu}}\nabla_{\ti{\epsilon}}\Phi
\ti{D}_{\bar{{\rm z}}}\ti{\phi}^{\ti{\nu}}_0
\ti{D}_{\rm z}\ti{\phi}^{\ti{\epsilon}}_0)
-2\ti{\delta} \ti{D}_{\bar{{\rm z}}}(\nabla_\mu\nabla_\rho \Phi
D_{\bar{{\rm z}}}\phi^\mu_0 D_{\rm z}\phi^\rho_0) \right\}
      $$
for the symmetry breaking terms.
If we set them to zero we obtain the following conditions on the manifold
\begin{equation} \ba{l}
 R_{{\ti{\nu}}\mu} -2\nabla_{\ti{\nu}}\nabla_\mu \Phi =0    \\
\nabla_\mu\nabla_\rho \Phi=0 ~~~~~
\nabla_{\ti{\nu}}\nabla_{\ti{\epsilon}} \Phi = 0 \\
- \frac{1}{4}  R + \nabla^2 \Phi - (\nabla \Phi)^2 = 0
\ea \label{eman} \end{equation}
The term proportional to the dimension  cancels with the contribution of
the corresponding ghosts for $D=2$ ($D=$ complex dimension) for the N=2 string
and it is $D=3$ in superstring compactifications. In this case
it is easy to see that the dilaton is set to a constant by the
equations (\ref{eman}).
By rearranging the equations (\ref{eman}) the condition $\nabla^2\Phi =
2(\nabla \Phi)^2$ is obtained. Integrating both sides and using that
we are in a compact euclidean space, we obtain
$\nabla \Phi = 0 $. In the
case of the N=2 string, it is not a new degree of freedom
(only zero momentum states are allowed).
This should be expected from the fact that the only physical state is the
vacuum $^{\cite{ademollo}}$, whose
expectation value is described by the K\"ahler potential.
If the dilaton is a constant we obtain the Ricci-flatness condition in
accordance with the $\beta$ function  method result
$^{\cite{alvgaumecounter}}$.
Note as well that the equations (\ref{eman})
are simply the ones obtained $^{\cite{callanbackground,gerardo}}$ for the N=1
sigma model in the case of a K\"ahler
metric, as is expected from the universality properties of supersymmetric
sigma models: the models with extended supersymmetry are obtained from the
N=1 model restricting the target manifold.

\section{conclusions}

The method we have developed maintains the N=2 supersymmetry and is partially
covariant. A reason  explaining why a completely covariant method is
not possible is given. Our method is effectively covariant for some
calculations, as
the ghosts can be absent. In fact, if this method is applied to the calculation
of the $\beta$ function, it attains its maximum simplicity because in that case
power counting arguments leave only vertices from $S_K$ (\ref{enserio}).
This is encouraging for undertaking higher loop computations.
Previous calculations
in this model were done either using a non covariant method $^{\cite{grisaru}}$
or a non
supersymmetric one $^{\cite{japoneses}}$.

The calculation of the supercurrent operator product expansion shows, on the
one hand, how the method works and, on the other, makes it clear how the
conformal anomaly breaks the algebra. The dilaton field was included and
 severe constraints on it were found, as  could be expected from the N=2 string
spectrum $^{\cite{ademollo}}$.

This method can also be extended to treat the models
considered in ref. \cite{rocek} which include  the antisymmetric tensor.

\vspace{1cm}

\noindent
Acknowledgments:

We thank F. Toppan for some important observations and
R. Trinchero for helpful conversations.


\newpage

\begin{center}
FIGURE CAPTIONS
\end{center}

\underline{Figure 1:}\\
Graphs of order zero in $\alpha'$ contributing
to $\langle JJ \rangle $. Only graphs not related by interchange of
${\cal Z} \leftrightarrow {\cal Z}'$ or by complex conjugation (obtained
reversing all lines) are shown.

{\footnotesize
{}~~~~~~~ stands for a $\langle \xi\ti{\xi}\rangle $ propagator\ \ \ \ \ \ \ \
\ \ \ \
stands for a $\langle \eta\ti{\xi}\rangle $ propagator \ \ \ \ \ \ \ \ \ \ \
stands for
a ghost propagator ~~~~~~~~~ stands for a background field  \ \ \ $\times$
stands for a supercurrent \ \ \ $\otimes$ stands for the dilaton supercurrent
 \ \ \ \
R stands for a curvature \ \ \ $\omega$ stands for a connection.}

\underline{Figure 2:}\\
Graphs of order zero in $\alpha'$ contributing to
$\langle J \rangle $.

\underline{Figure 3:}  \\
Graphs of order one in $\alpha'$ contributing to
 $\langle JJ \rangle $.


\begin{thebibliography}{99}
\bibitem{vafa} H. Ooguri and C.Vafa Princeton preprint HUTP-90/A024 and
HUTP-91/A003.

\bibitem{todosn=2} P. Candelas, G. Horowitz, A. Strominger and E. Witten,
Nucl.Phys. B258 (1985)~46,

 W. Boucher,  D.  Frie\-dan, A. Kent, Phys.Lett. 172B(1986) 316,

 A. Sen, Nucl.Phys. B278(1986) 289; Nucl.Phys. B284(1987) 423,

 T. Banks, L. J. Dixon, D. Friedan, E. Martinec, Nucl.Phys. B299(1988) 613,

 D. Gepner ``Lectures on N=2 String Theories''
Summer School in High Energy Physics and Cosmology, Trieste, (1989).

\bibitem{alvgaumen=2} L. Alvarez-Gaume and D. Freedman, Commun. Math. Phys.
80, (1981) 443.

\bibitem{zumino} B. Zumino Phys. Lett. 87b (1979) 203.

\bibitem{honerkamp} J. Honerkamp Nucl. Phys. B36 (1972) 130.

\bibitem{alvgaumecoord} L. Alvarez-Gaume, D. Feedman and S. Mukhi, Ann. Phys.
134 (1981) 85.

\bibitem{howe} P. Howe G. Papadopoluos and K. Stelle Phys. Lett. B174 (1986)
405.

\bibitem{grisaru} M.T.Grisaru, A.E.M.Van de Ven and D.Zanon, Phys. Lett. B 173
(1986) 423.

\bibitem{callanbackground} C. Callan, D. Friedan, E. Martinec and M. Perry,
Nuc. Phys.
B262 (1985) 593.

\bibitem{banks} T. Banks, D. Nemeschansky and A. Sen Nucl. Phys. B277 (1986)
67.

\bibitem{gerardo} G. Aldazabal F. Hussain and R. Zhang, Phys. Lett. 185b (1987)
89; ``Superconformal invariance and superstrings in background fields'' ICTP
preprint, IC/86/400.

\bibitem{japoneses} T. Itoh and M. Takao, Int. J. Mod. Phys. A12 (1990) 2265.

\bibitem{divecchia} P. Di Vecchia, J. Petersen and H. Zheng, Phys. Lett. 162B
(1985) 327.

\bibitem{ademollo} M. Ademollo, L. Brink, A. D'Adda, R. D'Auria, E. Napolitano,
S. Sciuto,
 E. Del Giudice, P. Di Vecchia, S. Ferrara, F. Gliozzi, R. Musto, R. Pettorino
and J.H. Schwarz, Phys. Lett. B271 (86)93 , Nucl. Phys. B111 (1976) 77.

\bibitem{alvgaumecounter} L. Alvarez-Gaume and D.Z. Freedman, Phys. Rev. D22
(1980) 846.

\bibitem{rocek} S.J. Gates, C.M. Hull and M. Ro\u{c}ek, Nucl. Phys B248 (1984)
157.

\end{thebibliography}
\end{document}